\documentclass[prd,twocolumn,showpacs]{revtex4}
\usepackage{epsfig}

\begin{document}

\title{Ideal hydrodynamics inside as well as outside non-rotating black hole:
Hamiltonian description in the Painlev{\'e}-Gullstrand coordinates}
\author{V. P. Ruban}
\email{ruban@itp.ac.ru}
\affiliation{Landau Institute for Theoretical Physics RAS, Moscow, Russia} 

\date{\today}

\begin{abstract}
It is demonstrated that with using Painlev{\'e}-Gullstrand coordinates in their quasi-Cartesian 
variant, the Hamiltonian functional for relativistic perfect fluid hydrodynamics near 
a non-rotating black hole differs from the corresponding flat-spacetime Hamiltonian 
just by a simple term.
Moreover, the internal region of the black hole is then described uniformly together with 
the external region, because in Painlev{\'e}-Gullstrand coordinates there is no singularity 
at the event horizon.
An exact solution is presented which describes stationary accretion of an ultra-hard matter
($\varepsilon\propto n^2$) onto a moving black hole until reaching the central singularity.
Equation of motion for a thin vortex filament on such accretion background is derived 
in the local induction approximation. The Hamiltonian for a fluid having ultra-relativistic
equation of state $\varepsilon\propto n^{4/3}$ is calculated in explicit form, 
and the problem of centrally-symmetric stationary flow of such matter is solved analytically.
\end{abstract}



\maketitle

\section{Introduction}

General-relativistic hydrodynamics is an important part of the modern astrophysics
(see, e.g., \cite{AC2007,Font2008}, and references therein). In some astrophysical problems 
it is possible to neglect thermal radiation and dissipative effects.
It such cases the model of perfect fluid is a good approximation. 
It is a well-known fact that equations of motion for a fluid in a curved spacetime 
can be derived with the help of simple physical arguments about the structure of the 
stress-energy tensor \cite{LL6}. Such an approach is especially productive for
numerical modeling (see, e.g., \cite{Font2008,num1,num2,num3,num4,num5}, and references therein).
But equations of ideal relativistic hydrodynamics also admit various variational formulations,
which fact is important for analytical investigations 
(see \cite{Taub1954,Schutz1970,Ray1972,Brown,CL1994}, and references therein).
In most variational approaches the method of a Lagrangian with kinematic constraints is used.
Therefore it is appropriate to make two observations. First, introduction of additional fields
-- Lagrangian multipliers -- increases the number of unknown functions over a
minimally required set, ant that is unwanted due to many reasons. Second, not all spacetime
coordinate systems are equally useful to study fluid dynamics. The second observation
is especially relevant in that cases when, despite matter motion, the gravitational field 
remains steady with a good accuracy (for instance, one can imply a fluid dynamics near a 
sufficiently massive black hole, as in the present work, or slow flows on a spatially nonuniform
static matter distribution background, as in Refs.\cite{R2000PRD,R2014Pisma}).
The simpler form a given metrics $ds^2=g_{ik}(t,{\bf r})dx^i dx^k$ has, the more compact
equations of fluid motion are, and the more chances exist to advance in understanding the physics 
of phenomena. It is also very important that components of the metric tensor should not have
non-physical singularities which are removable by a coordinate transformation \cite{PF1998}.

It follows from the above observations that for solving problems of general-relativistic
hydrodynamics it makes sense to choose the most suitable coordinate systems and also restrict
consideration to a minimally required set of unknown functions. The purpose of this work
is to demonstrate advantages of so called Painlev{\'e}-Gullstrand coordinates \cite{P1921,G1922}
in description of accretion flows of a perfect fluid inside as well as outside a non-rotating
black hole. These coordinates and their generalizations were successfully applied to
a number of physical problems \cite{PW2000,V2001,ZK2009,KSH2011}, but, to the best present 
author's knowledge, they were never previously used for consideration of three-dimensional 
flows of a fluid with arbitrary equation of state. It will be shown here that equations of
relativistic hydrodynamics in Painlev{\'e}-Gullstrand coordinates appear just a little more 
complex in comparison with the flat spacetime. More precisely, the Hamiltonian functional
of a hydrodynamical system differs by a single and simple term from the corresponding 
Hamiltonian in the Minkowskii space. Let us remind that the widely known Schwarzschild
metrics \cite{LL2}, which describes a black hole of mass $M$, by the change of the time coordinate
\begin{equation}\label{T_r}
t_{\rm S}=t-2\sqrt{2Mr}-2M\ln\frac{1-\sqrt{2M/r}}{1+\sqrt{2M/r}} 
\end{equation}
(see, for instance, \cite{PW2000}) is brought to the Painlev{\'e}-Gullstrand metrics:
\begin{equation}\label{P-G}
ds^2=dt^2 -\Big(dr+\sqrt{\frac{2M}{r}}\, dt\Big)^2 
-r^2(d\theta^2+\sin^2\theta \,\,d\phi^2).
\end{equation}
Here the geometrical units are in use, where the speed of light $c=1$, and the Newton constant $G=1$.
It is very essential that now a singularity at the event horizon $r_g=2M$ is absent, and this
feature of Painlev{\'e}-Gullstrand coordinates allows us to consider the entire space uniformly,
contrary to the Schwarzschild coordinates. If instead of the ``spherical'' spatial coordinates
one introduces in a standard way the ``Cartesian'' coordinates  ${\bf r}=(x,y,z)$, then the given
stationary metrics is rewritten in a remarkably compact form:
\begin{equation} \label{P-G-Cartesian}
ds^2=dt^2-(d{\bf r}-{\bf U} dt)^2,\qquad 
{\bf U}({\bf r})=-\frac{\bf r}{r}\sqrt{\frac{2M}{r}},
\end{equation}
and the scalar square here is in the simplest sense, that is the sum of squares of three
components. As we shall see later, another important advantage of metrics (\ref{P-G-Cartesian})
is its constant determinant $g=\mbox{det}\|g_{ik}\| =-1$.

\section{General structure of equations}

It is possible to avoid the redundant description of system with Lagrangian multipliers 
--- by making use of the generalized Euler equation concept \cite{R2000PRD,R2001PRE}.
The equation is applicable to a wide class of models, relativistic among others.
In this approach non-magnetized isentropic flows are described by two fields -- 
by the field of relative (``coordinate-'') density $\rho(t,{\bf r})$, and by the  
three-dimensional velocity field ${\bf v}(t,{\bf r})$. Though in the general case coordinates 
$(t,{\bf r})$ are arbitrary curvilinear, dynamics of the ``density''  
$\rho(t,{\bf r})$ obeys the continuity equation in its standard ``Cartesian'' form,
\begin{equation} \label{rho_t}
\rho_t+\nabla\cdot(\rho{\bf v})=0.
\end{equation}
The origin of this equation is purely kinematic. In the general relativity it corresponds to
the condition of zero divergence for the current 4-vector $n^i=ndx^i/ds$ \cite{LL6,LL2},
where the scalar $n$ is the (physical) density of number of conserved particles in the proper 
frame of reference, that is
\begin{equation}\label{n_conserv}
n^i_{;i}\equiv\frac{1}{\sqrt{-g}}\frac{\partial}{\partial x^i}
\left(\sqrt{-g}n\frac{dx^i}{d s}\right)=0.
\end{equation} 
From comparison Eq.(\ref{rho_t}) to Eq.(\ref{n_conserv}) one obtains the relation between
the field $n$ and the dynamical variables  $\rho$ and ${\bf v}$:
\begin{equation}\label{n_rho_v}
n=\frac{\rho}{\sqrt{-g}}
\sqrt{g_{00}+2g_{0\alpha}v^\alpha+g_{\alpha\beta}v^\alpha v^\beta},
\end{equation}
where  $\alpha$ and $\beta$  are three-dimensional (3D) tensorial indices.

It will be convenient for us to deal with the ``current density'' field 
${\bf j}\equiv \rho{\bf v}$, so that the continuity equation now is
\begin{equation} \label{rho_t_j}
\rho_t+\nabla\cdot{\bf j}=0,
\end{equation}
and the relation (\ref{n_rho_v}) is rewritten in the form
\begin{equation}\label{n_rho_j}
n=\sqrt{g_{00}\rho^2+2g_{0\alpha}\rho j^\alpha+g_{\alpha\beta}j^\alpha j^\beta}/\sqrt{-g}.
\end{equation}

Besides the kinematic equation (\ref{rho_t_j}), there is the second, dynamical, equation of motion,
which depends on the model under consideration. It is determined by some Lagrangian functional
${\cal L}\{\rho,{\bf j}\}$ and possesses the following structure (the generalized Euler equation;
see details in \cite{R2000PRD,R2001PRE}, where it was presented in terms of $\rho$ and ${\bf v}$):
\begin{equation}
\frac{\partial}{\partial t}\left(\frac{\delta {\cal L}}{\delta{\bf j}}\right)=
\left[\frac{{\bf j}}{\rho}\times
\mbox{curl}\left(\frac{\delta {\cal L}}{\delta{\bf j}}\right)
\right]+{\bf \nabla}\left(\frac{\delta {\cal L}}{\delta \rho}\right),
\label{dynequation}
\end{equation}
with 3D vector operators acting here in the same manner as in the usual Cartesian coordinates.
Eq.(\ref{dynequation}) is nothing else but the expressed in terms of $\rho$ and ${\bf j}$
variational Euler-Lagrange equation
\begin{equation}
\frac{\delta {\cal L}}{\delta{\bf x}({\bf a})}
-\frac{d}{dt}\frac{\delta {\cal L}}{\delta\dot{\bf x}(\bf a)}=0
\end{equation}
for the mapping ${\bf r}={\bf x}(t,{\bf a})$ which describes trajectory of each element of the 
liquid medium, labeled by a label ${\bf a}=(a_1,a_2,a_3)$. The vector
\begin{equation}\label{p_def}
{\bf p}=\delta {\cal L}/\delta{\bf j}=\delta {\cal L}/\delta\dot{\bf x}(\bf a)
\end{equation} 
is the canonical momentum of the liquid element. It is significant that the generalized (3D)
vorticity field ${\bf\Omega}=\mbox{curl}\,{\bf p}$ is ``frozen-in'' into the fluid, because it
obeys the equation
\begin{equation} 
{\bf\Omega}_t=\mbox{curl}[{\bf v}\times{\bf\Omega}],
\end{equation}
which follows from Eq.(\ref{dynequation}). In particular, there exists the class of purely 
potential flows, where ${\bf p}=\nabla\varphi$.

As we shall see later, after resolving with respect to temporal derivative ${\bf j}_t$, 
Eq.(\ref{dynequation}) acquires rather cumbersome form even in the flat spacetime.
Besides that, the variable ${\bf j}$ gives no advantages in description of potential flows.
Therefore it makes sense to consider also different equivalent systems of equations,
taking as the basic dynamical variables not $(\rho,{\bf j})$, but $(\rho,{\bf p})$ 
or $(n,{\bf p})$.

From the methodical point of view, the most preferable is the pair $(\rho,{\bf p})$, because
in that case the system of equations (\ref{rho_t_j}) and (\ref{dynequation}) acquires
the non-canonical Hamiltonian structure \cite{R2001PRE}:
\begin{eqnarray}
&&\rho_t=-\nabla\cdot\Big( \frac{\delta {\cal H}}{\delta{\bf p}}\Big),
\label{rho_t_Ham_noncanon_compress}\\
&&{\bf p}_t=
\left[\frac{1}{\rho}\Big( \frac{\delta {\cal H}}{\delta{\bf p}}\Big) 
\times\mbox{curl\,}{\bf p}\right]-\nabla\frac{\delta {\cal H}}{\delta \rho}.
\label{p_t_Ham_noncanon_compress}
\end{eqnarray}
The Hamiltonian ${\cal H}\{\rho,{\bf p}\}$ is the Legendre transform of the Lagrangian
on the vector variable ${\bf j}$: 
\begin{equation}
{\cal H}\{\rho,{\bf p}\}=\int ({\bf j}\cdot{\bf p})d{\bf r}-{\cal L},
\end{equation}
and instead of ${\bf j}$ here the solution of equation
${\bf p}=\delta{\cal L}\{\rho,{\bf j}\}/\delta{\bf j}$ should be substituted.
Let us note by the way that potential flows are described by the single pair of unknown functions
$\rho$ and $\varphi$, which are canonically conjugate in this case. A more general class
of flows can be parametrized by two pairs of canonically conjugate quantities,
$(\rho,\varphi)$ and $(\lambda,\mu)$, if the Clebsch representation is used,
\begin{equation}
{\bf p}=\nabla\varphi+\frac{\lambda}{\rho}\nabla\mu.
\end{equation}
Unfortunately, as we shall see later, it is only possible with special fluid equations of state
to resolve in a closed form the relation (\ref{p_def}) with respect to ${\bf j}$ and thus
perform the Legengre transform from the Lagrangian to the Hamiltonian.
Because of this reason, equations of motion in variables $(n,{\bf p})$ will be presented as well,
though seemingly not possessing a simple Hamiltonian structure, but rather elegant by themselves.

\section{Equations of hydrodynamics in metrics (\ref{P-G-Cartesian})}

The Lagrangian of relativistic hydrodynamics is determined by an equation of state $\varepsilon(n)$ 
of the fluid which is the relation between $n$ and the proper mass-energy density
(with a constant entropy per conserved particle). A general form of the Lagrangian 
${\cal L}\{\rho,{\bf j}\}$ is given below:
\begin{equation} \label{L_GR}
{\cal L}\!=\!-\!\int\!\sqrt{-g}\varepsilon\Big(
\sqrt{g_{00}\rho^2+2g_{0\alpha}\rho j^\alpha+g_{\alpha\beta}j^\alpha j^\beta}/\sqrt{-g}
\Big)d{\bf r}.
\end{equation}
Here, instead of $n$, the expression (\ref{n_rho_j}) has been substituted into equation of state.
The action functional $I=\int {\cal L}dt$ is a relativistic invariant, as it should be.
This Lagrangian is in accord with the well known stress-energy tensor of a perfect 
fluid \cite{LL6,LL2}. Physically acceptable functions $\varepsilon(n)$ should satisfy definite
requirements, which are not discussed here. In a number of previously suggested variational
formulations \cite{Taub1954,Schutz1970,Ray1972}, based on different sets of dynamical variables,
Ray's formulation \cite{Ray1972} is the most close to Lagrangian (\ref{L_GR}).
In the flat spacetime, and at small velocities, the integral (\ref{L_GR}) reduces to the customary 
difference between the kinetic and potential energies (with taking into account the rest energy).

Let us now make use of the metrics (\ref{P-G-Cartesian}), which allows us to write the Lagrangian
in the remarkably compact form:
\begin{equation} \label{L_P-G}
{\cal L}=-\int\varepsilon\Big(\sqrt{\rho^2-({\bf j}-{\bf U}\rho)^2}\Big)d{\bf r}.
\end{equation}
Only the presence of central-symmetric field ${\bf U}({\bf r})$ distinguishes this Lagrangian
from the Lagrangian in the flat spacetime. The corresponding variational derivatives are
\begin{eqnarray}
{\delta {\cal L}}/{\delta{\bf j}}&=&{\bf J}f(n),
\label{L_j}
\\
{\delta {\cal L}}/{\delta\rho}&=&-(\rho+{\bf U}\cdot{\bf J})f(n),
\label{L_rho}
\end{eqnarray}
where it has been denoted for brevity
\begin{equation}\label{relations}
{\bf J}={\bf j}-{\bf U}\rho,\quad n=\sqrt{\rho^2-{\bf J}^2},\quad f(n)=\varepsilon'(n)/n.
\end{equation}
With using the continuity equation, Eq.(\ref{dynequation}) can be represented in the form
\begin{equation}\label{dyn_PG}
{\bf J}_t f(n)-{\bf J}[\rho(\nabla\cdot{\bf j})+{\bf J}\cdot{\bf J}_t]f'(n)/n
={\bf R}f(n),
\end{equation}
where ${\bf R}f(n)$ is the right hand side:
\begin{equation}
{\bf R}f=\left[({\bf j}/\rho)\times\mbox{curl}({\bf J}f)\right]
-{\bf \nabla}[\left(\rho+{\bf U}\cdot{\bf J}\right)f].
\end{equation}
It follows from (\ref{dyn_PG}) that
\begin{equation}
{\bf J}\cdot{\bf J}_t=
\frac{{\bf R}\cdot{\bf J}f(n)+{\bf J}^2\rho(\nabla\cdot{\bf j})f'(n)/n}
{f(n)-{\bf J}^2f'(n)/n},
\end{equation}
and this relation allows us to resolve Eq.(\ref{dyn_PG}) with respect to time derivative
${\bf J}_t={\bf j}_t+{\bf U}(\nabla\cdot{\bf j})$
and thus represent the dynamical equation in the evolutionary form:
\begin{equation}\label{j_t}
{\bf j}_t=-{\bf U}(\nabla\cdot{\bf j})+{\bf R}
+{\bf J}\left[
\frac{\rho(\nabla\cdot{\bf j)}+{\bf R}\cdot{\bf J}}
{f(n)-{\bf J}^2f'(n)/n}\right]\frac{f'(n)}{n}.
\end{equation}
It is important that for physically acceptable equations of state $f(n)>0$, $f'(n)<0$,
and therefore the denominator in the above equation never takes zero value.

One cannot say that the  right hand side of Eq.(\ref{j_t}) is simple, but its unhandiness
is not connected with the spacetime curvature, but it is peculiar to relativistic hydrodynamics 
as such. More elegant is description in terms of fields $n$ and ${\bf p}$, although the continuity
equation looks more complicated here:
\begin{equation}
\frac{\partial}{\partial t}\Big(\frac{n}{w}\sqrt{w^2+{\bf p}^2}\Big)
+\nabla\cdot\Big(\frac{n}{w}\Big({\bf p}+{\bf U}\sqrt{w^2+{\bf p}^2}\Big)\Big)=0,
\label{contin_np}
\end{equation}
where $w=\varepsilon'(n)$ is the relativistic enthalpy per particle. The dynamic equation has
the following form:
\begin{eqnarray}
{\bf p}_t&=&\Big[\Big({\bf U}+\frac{\bf p}{\sqrt{w^2+{\bf p}^2}}\Big)
\times\mbox{curl}\,{\bf p}\Big]\nonumber\\
&&-\nabla\Big(\sqrt{w^2+{\bf p}^2}+{\bf U}\cdot{\bf p}\Big).
\label{dyn_np}
\end{eqnarray}
Eqs.(\ref{contin_np}) and (\ref{dyn_np}) are easily derived using formulas
(\ref{L_j}), (\ref{L_rho}), and (\ref{relations}). Obviously, with the help of Eq.(\ref{dyn_np}) 
it is possible to express in Eq.(\ref{contin_np}) the temporal derivative $n_t$ through 
$n$, ${\bf p}$, and their gradients, thus presenting the whole system in evolutionary form.
It may appear more convenient to take the enthalpy $w$ as a basic dynamical variable, and imply
that $n=n(w)=P'(w)$, where $P(w)$ is the pressure as a function of enthalpy (at fixed entropy).
Let us also note that in system (\ref{contin_np})-(\ref{dyn_np}) it is very easy to pass to
consideration of potential flows: it is sufficient to put ${\bf p}=\nabla\varphi$ and
remove the gradient operator in Eq.(\ref{dyn_np}). By acting in this way, we obtain that
\begin{equation}
w=\sqrt{(\varphi_t +{\bf U}\cdot\nabla\varphi)^2-(\nabla\varphi)^2}.
\end{equation}
It is interesting that now the continuity equation transforms from being kinematic to dynamic:
\begin{eqnarray}
&&-\frac{\partial}{\partial t}\Big[(\varphi_t +{\bf U}\cdot\nabla\varphi)
P'(w)/w\Big]
\nonumber\\
&&+\nabla\cdot\Big\{[\nabla\varphi -{\bf U}(\varphi_t +{\bf U}\cdot\nabla\varphi)]
P'(w)/w\Big\}=0.
\label{phi_eq}
\end{eqnarray}
The variational nature of this equation is evident. The corresponding action functional is
\begin{equation}\label{Schutz_pot}
I\{\varphi\}=\int 
P\Big(\sqrt{(\varphi_t +{\bf U}\cdot\nabla\varphi)^2-(\nabla\varphi)^2}\Big)d{\bf r}dt.
\end{equation}
The above expression is a particular case of the variational formulation by Schutz 
\cite{Schutz1970}; for more general form (in isentropic case) one has to change 
$\varphi_t\to \varphi_t+\tilde\lambda\mu_t$, 
$\nabla\varphi\to \nabla\varphi+\tilde\lambda\nabla\mu$ in integral (\ref{Schutz_pot}).
Hamiltonian formalism for such action was developed in Ref.\cite{DM1980}.
However, the substitution ${\bf p}=\nabla\varphi+\tilde\lambda\nabla\mu$ is able to describe 
not arbitrary flows, but only flows having trivial topology of the (frozen-in) vorticity,
because in that case ${\bf\Omega}=[\nabla\tilde\lambda\times\nabla\mu]$, and  vortex lines
at every time moment are intersections of two surface families $\tilde\lambda=const$, $\mu=const$.
The question about Hamiltonian structure of Eqs.(\ref{contin_np}) and (\ref{dyn_np}) directly,
with arbitrary topology of vortex lines, is not clear to the present author at the moment.

\section{Properties of the Hamiltonian}

Let us return to the problem of Hamiltonian description of flows in variables $(\rho,{\bf p})$.
If we succeed in analytical solution with respect to $J$ of the scalar equation
\begin{equation}\label{eq_p_J}
|{\bf p}|= Jf\Big(\sqrt{\rho^2- J^2}\Big),
\end{equation}
then it is possible to calculate the Hamiltonian explicitly. In our case we have
\begin{eqnarray}
{\cal H}\{\rho,{\bf p}\}=\int ({\bf j}\cdot{\bf p})d{\bf r}-{\cal L}
=\int \rho({\bf U}\cdot{\bf p})d{\bf r} &&
\nonumber\\
+ \int\Big[ \frac{J^2}{\sqrt{\rho^2-J^2}}
\varepsilon'\Big(\sqrt{\rho^2- J^2}\Big)+
\varepsilon\Big(\sqrt{\rho^2-J^2}\Big)\Big]d{\bf r},&&
\end{eqnarray}
and in the last integral everywhere instead of $J$ the solution of Eq.(\ref{eq_p_J}) 
should be substituted. It is a remarkable fact that the indicated integral is the 
Legendre transform of the Lagrangian for hydrodynamics in the Minkowskii spacetime.
Thus, we have obtained an interesting result: with using metrics (\ref{P-G-Cartesian}),
the Hamiltonian of a fluid in the presence of a non-rotating black hole is produced
by adding the term $\int \rho({\bf U}\cdot{\bf p})d{\bf r}$ to the Hamiltonian of the 
same fluid in flat spacetime:
\begin{equation}\label{H_PG}
{\cal H}\{\rho,{\bf p}\}=\int[\rho({\bf U}\cdot{\bf p})+H(\rho,|{\bf p}|)]d{\bf r}.
\end{equation}
The Hamiltonian, contrary to the action functional, is not relativistic invariant,
and the structural simplicity of expression (\ref{H_PG}) is due to the good choice of
coordinate system. The equations of motion take the form
\begin{eqnarray}
\rho_t&=&-\nabla\cdot\Big( \rho{\bf U}+ H_{\bf p}\Big),
\label{rho_t_U}\\
{\bf p}_t&=&
\left[\Big( {\bf U}+ H_{\bf p}/\rho\Big) 
\times\mbox{curl\,}{\bf p}\right]-\nabla(H_\rho+{\bf U}\cdot{\bf p}),
\label{p_t_U}
\end{eqnarray}
where $H_{\bf p}=({\bf p}/|{\bf p}|) H_{|{\bf p}|}$ 
(subscripts denote partial derivatives).

It is clear that function $H(\rho,|{\bf p}|)$ cannot be arbitrary, namely because its 
origin from a member of the specific family of Lagrangians depending on the combination 
$\sqrt{\rho^2-J^2}$. It is not difficult to show that each Hamiltonian of a fluid in 
Minkowskii spacetime should satisfy the simple first-order partial differential equation:
\begin{equation}\label{H_eq}
H_\rho H_{|{\bf p}|}=\rho |{\bf p}|.
\end{equation}
Indeed, by applying Legendre transform to the Hamiltonian, we obtain the Lagrangian, that is
\begin{equation}
|{\bf p}| H_{|{\bf p}|}-H=-\varepsilon\Big(\sqrt{\rho^2-H_{|{\bf p}|}^2}\Big).
\end{equation}
Next we consider two partial first-order derivatives of this equations, and with their help
we exclude $\varepsilon'$. After simplification and separation of multiplier 
$H_{|{\bf p}| |{\bf p}|}$ we arrive at Eq.(\ref{H_eq}). 
The complete integral of Eq.(\ref{H_eq}) is
\begin{equation}
H=C_1\rho^2/2 +{\bf p}^2/(2C_1) +C_2.
\end{equation}
Solution, corresponding to some physically acceptable equation of state, is obtained
in accordance with the general rules by excluding $C_1$ from the algebraic system of equations
\begin{equation}
H=C_1\rho^2/2 +{\bf p}^2/(2C_1) + F(C_1),
\end{equation}
\begin{equation}
\rho^2/2 -{\bf p}^2/(2C_1^2) + F'(C_1)=0,
\end{equation}
with appropriate function $F(C_1)$ \cite{ZP2003}. Similarly to the direct calculation 
of the Hamiltonian through the Legendre transform, the problem reduces to solution of an algebraic 
equation. Below, two examples of equation of state will be considered, when the Hamiltonian 
is found in explicit form.

\section{Stiff matter}

In a number of works, the ultra-hard equation of state was considered
$\varepsilon=n^2/2$ (stiff matter; see, e.g., \cite{hard1,hard2,hard3}, and references therein).
In this case  $w=n$, and the Hamiltonian $H=(\rho^2+{\bf p}^2)/2$. What is essential,
Eq.(\ref{phi_eq}) for potential flows becomes strictly linear and it coincides with equation 
for a massless scalar field. In particular, formation of shock waves is not possible within 
this model. It is not difficult to find solutions of Eq.(\ref{phi_eq}) describing stationary
accretion in the presence of a uniform matter flow at the infinity:
\begin{eqnarray}
\varphi_{\rm a}&=&\rho_\infty\Big[-t +2\sqrt{2Mr}
-4M\ln\Big(1+\sqrt{2M/r}\Big)\Big]\nonumber\\
&& +(1-M/r)({\bf r}\cdot{\bf p}_\infty),
\label{hard_solution}
\end{eqnarray}
where $\rho_\infty=\sqrt{w_\infty^2+({\bf p}_\infty)^2}$. Let us note that this expression
has no singularity at the gravitational radius $r_g=2M$, contrary to the same solution but
calculated in the Schwarz\-schild coordinates \cite{hard1}. The singularity is canceled
namely due to the change of time coordinate (\ref{T_r}).

As to vortex flows of stiff matter, they are described by a system which is not linear:
\begin{eqnarray}
&&\rho_t=-\nabla\cdot\Big( \rho{\bf U}+ {\bf p}\Big),
\label{rho_t_hard}\\
&&{\bf p}_t=
\left[\Big( {\bf U}+ {\bf p}/\rho\Big) 
\times\mbox{curl\,}{\bf p}\right]-\nabla(\rho+{\bf U}\cdot{\bf p}).
\label{p_t_hard}
\end{eqnarray}
If we consider the dynamics of relatively weak vortex disturbances at the background
flow determined by potential (\ref{hard_solution}), then equation for vorticity,
neglecting density perturbations, can be written as follows:
\begin{equation}\label{Omega_t_hard}
{\bf \Omega}_t=\mbox{curl\,}\Big[\Big({\bf U}+\frac{\nabla\varphi_{\rm a}
+\mbox{curl}^{-1}{\bf \Omega}}
{(\rho_\infty-{\bf U}\cdot\nabla\varphi_{\rm a})}\Big)\times{\bf \Omega}\Big],
\end{equation}
where the gradient of the stationary potential is given by the following formula
(for simplicity, we have normalized length scale to the gravitational radius $r_g$):
\begin{eqnarray}
\nabla\varphi_{\rm a}&=&\rho_\infty\frac{\bf r}{\sqrt{r^3}}
\Big(1+\frac{1}{r+\sqrt{r}}\Big)\nonumber\\
&&+{\bf p}_\infty\Big(1-\frac{1}{2r}\Big) +\frac{\bf r}{2r^3}({\bf r}\cdot{\bf p}_\infty).
\label{p_accr_hard}
\end{eqnarray}
Within Eq.(\ref{Omega_t_hard}) one can investigate the motion of frozen-in vortex structures,
in the same manner as it was done in works \cite{R2000PRD,R2001PRE}.
Let us designate for brevity 
\begin{eqnarray}
\bar\rho({\bf r})&=&\rho_\infty-{\bf U}\cdot\nabla\varphi_{\rm a}
\nonumber\\
&=&\rho_\infty\Big[1+\frac{1}{r}\Big(1+\frac{1}{r+\sqrt{r}}\Big)\Big]
+\frac{({\bf r}\cdot{\bf p}_\infty)}{\sqrt{r^3}},
\label{rho_accr_hard}
\end{eqnarray}
\begin{eqnarray}
{\bf S}({\bf r})&=&\bar\rho{\bf U}+\nabla\varphi_{\rm a}\nonumber\\
&=&-\rho_\infty\frac{{\bf r}}{r^3}+{\bf p}_\infty\Big(1-\frac{1}{2r}\Big)
-\frac{{\bf r}}{2r^3}({\bf r}\cdot{\bf p}_\infty).
\label{S_accr_hard}
\end{eqnarray}
It is easy to see that the stream ${\bf S}({\bf r})$ is solenoidal. There exists its
vector potential, which can be represented as ${\bf\Psi}={\bf e}_{\phi}F(m,z)$, where
$m=(x^2+y^2)/2$, and ${\bf e}_{\phi}$ is the unit vector in the azimuthal direction,
with $z$ axis oriented along ${\bf p}_\infty$. An explicit expression is given below:
\begin{equation}\label{vec_pot_S_hard}
{\bf\Psi}=\frac{\rho_\infty{\bf e}_{\phi}z}{\sqrt{2m}\sqrt{2m+z^2}}
+|{\bf p}_\infty| {\bf e}_{\phi}\Big(\frac{1}{2}-\frac{1}{2\sqrt{2m+z^2}}\Big)\sqrt{2m}.
\end{equation}

Let us consider for example a thin closed vortex filament with circulation $\Gamma$.
Then it follows from Eq.(\ref{Omega_t_hard}) that in so called local induction approximation 
(LIA; see, e.g., \cite{R2000PRD,R2001PRE}, and references therein) the dynamics of the filament 
shape ${\bf R}(\xi,t)$ obeys the equation
\begin{equation}\label{LIA_hard}
[{\bf R}_\xi\times{\bf R}_t]\bar\rho({\bf R})
=[{\bf R}_\xi\times{\bf S}({\bf R})]
-\Lambda\partial_\xi({\bf R}_\xi/|{\bf R}_\xi|).
\end{equation}
Here $\xi$ is an arbitrary longitudinal parameter, $\Lambda=(\Gamma/4\pi)\ln(L/d)$ is
the local induction constant,  $L$ is a typical size of vortex filament, 
$d$ is its small width. This equation obeys variational principle with the Lagrangian
of the form \cite{R2000PRD,R2001PRE}
\begin{equation}\label{variat_LIA_hard}
{\cal L}_{\Gamma}=\Gamma\oint\big([{\bf R}_\xi\times{\bf R}_t]\cdot{\bf D}({\bf R})\big) d\xi 
-{\cal H}_{\Gamma}\{{\bf R}\},
\end{equation}
where the vector function {\bf D}({\bf r}) satisfies the condition 
\begin{equation}
\nabla\cdot{\bf D}({\bf r})=\bar\rho({\bf r}).
\end{equation}
The local induction Hamiltonian ${\cal H}_{\Gamma}\{{\bf R}\}$ in our case looks as follows:
\begin{equation}\label{H_LIA_hard}
{\cal H}_{\Gamma}\{{\bf R}\}/\Gamma=\oint{\bf\Psi}({\bf R})\cdot{\bf R}_\xi\, d\xi
+\Lambda\oint|{\bf R}_\xi| d\xi.
\end{equation}
In the simplest configuration the filament is a coaxial vortex ring with a radius $\sqrt{2m(t)}$
and a center position  $z(t)$. Phase trajectories of the vortex in $(z,m)$ plane are determined
by level contours of its Hamiltonian 
\begin{equation}\label{traject_LIA_hard}
H_\Lambda=\frac{\rho_\infty z}{\sqrt{2m+z^2}}
+|{\bf p}_\infty|\Big(m-\frac{m}{\sqrt{2m+z^2}}\Big)+\Lambda\sqrt{2m},
\end{equation}
which is easily calculated with the help of formulas (\ref{vec_pot_S_hard}) and (\ref{H_LIA_hard}).
Equations of motion are almost canonical:
\begin{equation}
\dot z\bar\rho(m,z)=\partial H_\Lambda/\partial m,\quad
-\dot m\bar\rho(m,z)=\partial H_\Lambda/\partial z.
\end{equation}
Note that depending on the vorticity direction, parameter $\Lambda$ can be positive or negative.
The requirement of weakness of the vortex is expressed by condition 
$|\Lambda|/\rho_\infty\lesssim 1$. 

In an analogous manner it is possible to study a system of several coaxial vortex rings
in the accretion stream ${\bf S}({\bf r})$ with the density profile $\bar\rho({\bf r})$.
In the Hamiltonian, it is then necessary to take into account interactions between the rings.
The corresponding terms are expressed through elliptic integrals, in the same way as in the 
usual incompressible hydrodynamics (see, e.g., \cite{rings,BKP2013}, and references therein).
We shall not dwell on this question.

Thus, the model of stiff matter is the most simple in relativistic hydrodynamics. But unfortunately,
the ultra-hard equation of state violates the principle of non-negativeness of the 
stress-energy tensor trace of a fluid medium \cite{LL2}. Ultra-relativistic equation of state, 
$\varepsilon\propto n^{4/3}$, seems more adequate for studying flows near a black hole
and especially inside it. Now we are passing to its consideration.

\section{Ultra-relativistic matter} 

Let us demonstrate that the Hamiltonian can be calculated in closed form with 
$\varepsilon=(3/4) n^{4/3}$ (the multiplier is not very important, it has been 
introduced for convenience). Indeed, we then have the relation 
$|{\bf p}|= J(\rho^2- J^2)^{-1/3}$, which is
reduced to the cubic equation
\begin{equation}\label{eq_p_J_ultra}
|{\bf p}|^3(\rho^2- J^2)= J^3.
\end{equation}
Solving this equation, we obtain that ${\bf J}=H_{\bf p}$, where
\begin{equation}\label{H_p_ultra}
H_{\bf p}=\frac{{\bf p}\rho^{2/3}}
{\Big(\frac{1}{2}+\sqrt{\frac{1}{4}-\frac{|{\bf p}|^6}{27\rho^2}}\Big)^{\frac{1}{3}}+
 \Big(\frac{1}{2}-\sqrt{\frac{1}{4}-\frac{|{\bf p}|^6}{27\rho^2}}\Big)^{\frac{1}{3}}},
\end{equation}
and in the case of imaginary values of the square root, solutions having minimal argument
should be chosen when complex cubic roots are computed. The sum of two expressions in the 
denominator always remains purely real. The function $H(\rho,|{\bf p}|)$ is given by the
following expression:
\begin{eqnarray}\label{H_ultra}
H=\frac{|{\bf p}|^2\rho^{2/3}}
{\Big(\frac{1}{2}+\sqrt{\frac{1}{4}-\frac{|{\bf p}|^6}{27\rho^2}}\Big)^{\frac{1}{3}}+
 \Big(\frac{1}{2}-\sqrt{\frac{1}{4}-\frac{|{\bf p}|^6}{27\rho^2}}\Big)^{\frac{1}{3}}}
&&\nonumber\\
+\frac{(3/4)\rho^{4/3}}
{\Big[ \Big(\frac{1}{2}+\sqrt{\frac{1}{4}-\frac{|{\bf p}|^6}{27\rho^2}}\Big)^{\frac{1}{3}}+
 \Big(\frac{1}{2}-\sqrt{\frac{1}{4}-\frac{|{\bf p}|^6}{27\rho^2}}\Big)^{\frac{1}{3}}\Big]^2}.&&
\end{eqnarray}
For calculation of the partial derivative $H_\rho$, one can use Eq.(\ref{H_eq}), which
gives, together with Eq.(\ref{H_p_ultra}), that
\begin{equation}\label{H_rho_ultra}
H_{\rho}=\rho^{\frac{1}{3}}
\Bigg[\Bigg(\frac{1}{2}+\sqrt{\frac{1}{4}-\frac{|{\bf p}|^6}{27\rho^2}}\Bigg)^{\frac{1}{3}}+
\Bigg(\frac{1}{2}-\sqrt{\frac{1}{4}-\frac{|{\bf p}|^6}{27\rho^2}}\Bigg)^{\frac{1}{3}}\Bigg].
\end{equation}
Expressions (\ref{H_p_ultra}) and (\ref{H_rho_ultra}), though not very simple, but are still
good to deal with them. They can be substituted into Eqs.(\ref{rho_t_U}) and (\ref{p_t_U}).

Inasmuch as matter cannot be at rest under the event horizon, the main hydrodynamic regime
near a black hole, with any equation of state, is an accretion of the fluid and its subsequent
falling onto the central singularity. Accretion was investigated in many works (see, e.g.,
\cite{PF1998,accr1,accr2,accr3,accr4}, and references therein). The most simple its kind for
non-rotating black hole is a central-symmetric stationary flow. In the Painlev{\'e}-Gullstrand
coordinates such flows are described by the system of two algebraic equations, which follow
from Eqs.(\ref{contin_np})-(\ref{dyn_np}) (here we have normalized the length scale to $r_g$,
and the enthalpy to its value at the infinity):
\begin{equation}
\frac{n(w)}{w}\Big(p-\sqrt{\frac{(w^2+p^2)}{r}}\Big)=-\frac{A}{r^2},
\end{equation}
\begin{equation}
\sqrt{w^2+p^2}-p/\sqrt{r}=1,
\end{equation}
where $p$ is the radial component of the canonical momentum (do not mix up with the pressure $P$!),
$A$ is a positive constant, which is however not arbitrary; it should result in a physically
acceptable solution (the curve should pass through the so called critical point 
\cite{accr1,accr2,accr3,accr4}). Expressing $w$ with the help of the second equation and
substituting it into the first one, we obtain a relation of the form  $r^2S(p,r)=-A$, which
determines the dependence $p(r)$ implicitly. For a stiff matter, the solution is contained in
Eq.(\ref{p_accr_hard}); as to an ultra-relativistic fluid, when $n(w)=w^3$, the system reduces to
the cubic equation for $p$:
\begin{equation}\label{A}
r^2[(1+p/\sqrt{r})^2-p^2][p(1-1/r)-1/\sqrt{r}]=-A.
\end{equation}
The critical (saddle) point is determined by the conditions of zero values for the first-order 
partial derivatives of the left hand side of Eq.(\ref{A}) on variables $r$ and $p$.
Its numerical parameters are the following:
$A=\sqrt{27/4}\approx 2.5981$, $p_{\rm cr}=\sqrt{6}-\sqrt{3}\approx 0.71744$, 
$r_{\rm cr}= 3/2$. 
An analytical formula for the solution $p(r)$ has the following form:
\begin{eqnarray}
&&
p(r)=\frac{\sqrt{r}}{r-1}\nonumber\\
&&+\frac{r}{\sqrt{3}(1-r)}
\Bigg[\Bigg(\frac{27(1-r)}{4r^3}+\sqrt{\frac{27^2(1-r)^2}{16r^6}-1}\Bigg)^{\frac{1}{3}}
 \nonumber\\    
&&\quad\qquad+\Bigg(\frac{27(1-r)}{4r^3}-\sqrt{\frac{27^2(1-r)^2}{16r^6}-1}\Bigg)^{\frac{1}{3}}
\Bigg],
\label{p_ultra}
\end{eqnarray}
and in the case of imaginary square roots that branch should be chosen from the three, 
which guarantees a smooth dependence $p(r)$  at the entire interval $r>0$.
With the indicated value  $A$, the level contours of the left hand side of Eq.(\ref{A})
in $(r,p)$ plane consist of three curves, two of them intersecting at the critical point.
One from the intersecting curves continues in a smooth manner under the event horizon $r=1$
[it is the physical solution, and $p(1)=(\sqrt{27/4}-1)/2\approx 0.8$], while the two other
curves go to infinity as $r\to 1$. What is interesting, as $r\to 0$, the physical solution 
does not diverge but it tends to a finite value $p(0)=(27/4)^{1/6}\approx 1.38$.

The stationary density profile  $\bar\rho(r)$ now is expressed by the formula
\begin{equation}\label{rho_ultra}
\bar\rho(r)=[(1+p(r)/\sqrt{r})^2-p^2(r)](1+p(r)/\sqrt{r}),
\end{equation}
where the dependence (\ref{p_ultra}) has to be substituted.

Relatively weak vortex disturbances, practically not affecting the density profile $\bar\rho$,
are described by the following system of equations:
\begin{equation}\label{Omega_t_ultra}
{\bf \Omega}_t=\mbox{curl\,}\Big[\Big(-A\frac{\bf r}{r^3}+H_{\bf pp}({\bf r})\tilde{\bf p}
\Big)\times\frac{\bf \Omega}{\bar\rho({\bf r})}\Big],
\end{equation}
\begin{equation}\label{contin_ultra}
\nabla\cdot(H_{\bf pp}\tilde{\bf p})=0,
\end{equation}
\begin{equation}\label{vort_ultra}
\mbox{curl\,}\tilde{\bf p}={\bf \Omega},
\end{equation}
where $H_{\bf pp}({\bf r})$ is the matrix of second derivatives of the Hamiltonian (\ref{H_ultra})
on the components of vector ${\bf p}$, evaluated at the stationary solution. A more detailed
investigation of system (\ref{Omega_t_ultra})-(\ref{vort_ultra}), including a derivation
of an equation of motion for a thin vortex filament, is planned on the future.

\section{Conclusion}

Thus, it has been shown in this work that use of the Painlev{\'e}-Gullstrand coordinates
really allows one to follow the fate of falling matter under the horizon of a black hole. 
For the simplest equation of state, stiff matter, we have succeeded in obtaining new results
about vortex flows. Probably, stiff matter admits an analogous consideration in the case
of rotating black hole as well, though the corresponding Kerr metrics is more complicated,
and its spatial part cannot be reduced to the flat form. As far as more realistic 
equations of state are concerned, in particular an ultra-relativistic fluid, there investigation
of vortex structure dynamics needs more serious computational efforts, but from general 
point of view the road is open.



\begin{thebibliography}{99}

\bibitem{AC2007} N. Andersson and G. L. Comer,
Living Rev. Relativity {\bf 10}, 1 (2007).

\bibitem{Font2008} J. A. Font,
Living Rev. Relativity {\bf 11}, 7 (2008).

\bibitem{LL6}  L. D. Landau and E. M. Lifshitz, {\it Fluid Mechanics}
(Pergamon Press, New York) [Russian original, Nauka, Moscow, 1988].

\bibitem{num1} M. D. Duez, P. Marronetti, S. L. Shapiro, and T. W. Baumgarte,
Phys. Rev. D {\bf 67}, 024004 (2003). 

\bibitem{num2} L. Baiotti, I. Hawke, P. J. Montero, {\it et al}.,
Phys. Rev. D {\bf 71}, 024035 (2005). 

\bibitem{num3} H. Dimmelmeier, J. Novak, J. A. Font, {\it et al}.,
Phys. Rev. D {\bf 71}, 064023 (2005). 

\bibitem{num4} M. D. Duez, Y. T. Liu, S. L. Shapiro, and B. C. Stephens,
Phys. Rev. D {\bf 72}, 024028 (2005).

\bibitem{num5} Z. B. Etienne, J. A. Faber, Y. T. Liu, {\it et al}.,
Phys. Rev. D {\bf 77}, 084002 (2008). 

\bibitem{Taub1954} A. H. Taub, Phys. Rev. {\bf 94}, 1468 (1954).

\bibitem{Schutz1970} B. F. Schutz, Jr.,  Phys. Rev. D {\bf 2}, 2762 (1970).

\bibitem{Ray1972} J. R. Ray, J. Math. Phys. {\bf 13}, 1451 (1972).

\bibitem{Brown} J. D. Brown, Class. Quantum Grav. {\bf 10}, 1579 (1993).

\bibitem{CL1994} G. L. Comer and D. Langlois, Class. Quantum Grav. {\bf 11}, 709 (1994).

\bibitem{R2000PRD} V. P. Ruban, Phys. Rev. D {\bf 62}, 127504 (2000).

\bibitem{R2014Pisma} V. P. Ruban, JETP Letters {\bf 99}, ??? (2014),
[Russian original: Pis'ma v ZhETF {\bf 99}, 141 (2014)].

\bibitem{PF1998} P. Papadopoulos and J. A. Font, Phys. Rev. D {\bf 58}, 024005 (1998). 

\bibitem{P1921} P. Painlev{\'e}, C. R. Acad. Sci. Paris {\bf 173}, 677 (1921).

\bibitem{G1922} A. Gullstrand, Arkiv. Mat. Astron. Fys. {\bf 16}, 1 (1922).

\bibitem{PW2000} M. K. Parikh and F. Wilczek, Phys. Rev. Lett. {\bf 85}, 5042 (2000).

\bibitem{V2001} G. E. Volovik, Pis'ma v ZhETF {\bf 73}, 721 (2001).

\bibitem{ZK2009} J. Ziprick and G. Kunstatter, Phys. Rev. D {\bf 79}, 101503(R) (2009).

\bibitem{KSH2011} Y. Kanai, M. Siino, and A. Hosoya, 
Progr. Theor. Phys. {\bf 125}, 1053 (2011).

\bibitem{LL2}  L. D. Landau and E. M. Lifshitz, 
{\it The Classical Theory of Fields} (Pergamon Press, Oxford, 1980).

\bibitem{R2001PRE} V. P. Ruban, Phys. Rev. E {\bf 64}, 036305 (2001).

\bibitem{DM1980} J. Demaret and V. Moncrief, Phys. Rev. D {\bf 21}, 2785 (1980).

\bibitem{ZP2003} see any textbook on first-order partial differential equations.

\bibitem{hard1} L. I. Petrich, S. L. Shapiro, and S. A. Teukolsky,
Phys. Rev. Lett. {\bf 60}, 1781 (1988).

\bibitem{hard2}  S. L. Shapiro, Phys. Rev. D {\bf 39}, 2839 (1989).

\bibitem{hard3} E. Babichev, S. Chernov, V. Dokuchaev, and Yu. Eroshenko,
Phys. Rev. D {\bf 78}, 104027 (2008).

\bibitem{rings} B. N. Shashikanth and J. E. Marsden, Fluid Dyn. Res. {\bf 33}, 333 (2003).

\bibitem{BKP2013} E. Yu. Bannikova, V. M. Kontorovich, and S. A. Poslavsky,
JETP {\bf 117}, 378 (2013).

\bibitem{accr1} F. C. Michel, Astrophys. Space Sci. {\bf 15}, 153 (1972).

\bibitem{accr2} V. S. Beskin and V.I. Paryev, Uspekhi Fiz. Nauk {\bf 163}, 95 (1993).

\bibitem{accr3} V. S. Beskin and Y. N. Pidoprygora,  Zh. Exp. Teor. Phys. {\bf 107}, 1025 (1995).
 
\bibitem{accr4} E. O. Babichev, V. I. Dokuchaev, and Y. N. Eroshenko, JETP {\bf 100}, 528 (2005).

\end{thebibliography}
\end{document}